\documentclass[11pt]{article}

\usepackage[T1]{fontenc}
\usepackage[utf8]{inputenc}
\usepackage{lmodern}

\usepackage{amsmath, amssymb, amsthm}

\usepackage{cite}

\usepackage{geometry}
\usepackage[hidelinks]{hyperref}

\usepackage{enumitem}

\theoremstyle{plain}
\newtheorem{theorem}{Theorem}[section]

\theoremstyle{definition}
\newtheorem{definition}[theorem]{Definition}

\theoremstyle{remark}

\title{Expressive Boundedness of Authoritative DNS Response Selection}
\author{
Chris Bertinato\thanks{
IBM.\\
\texttt{chris.bertinato@ibm.com}
}
}
\date{February 2026}

\begin{document}

\maketitle

\begin{abstract}
Authoritative Domain Name System (DNS) response-selection semantics determine query-time outcomes from resolver-visible query context, authoritative candidate data, and associated metadata. Although widely deployed as traffic steering, these semantics have not been formalized independently of particular configuration languages or implementations.

We show that authoritative DNS response selection inhabits a bounded semantic domain induced by protocol constraints, including finiteness, RRset atomicity, totality, termination, cacheability, and restriction to resolver-visible query context and materialized candidate data and metadata available at evaluation time. Together, these constraints determine the admissible input domain, outcome space, and function class. We formalize response selection as DNS-admissible functions over finite structured candidate domains comprising individual answers and finite answer groupings. Every such function admits a finite compositional normal form that distinguishes observationally relevant combinations of query context and metadata, then restricts and selects over the corresponding candidate domain. Here, "normal form" asserts existence, not uniqueness, canonicalization, or minimality.

These normal forms expose an intrinsic compositional structure with a natural semiring interpretation. This supports principled reasoning about equivalence, expressiveness, approximation, semantic collapse, representability, and portability. Authoritative systems, configuration models, and serialized encodings are modeled uniformly as semantic restrictions of the shared bounded domain, grounded in protocol semantics rather than implementation detail.
\end{abstract}

\section{Introduction}
Authoritative Domain Name System (DNS) response-selection semantics determine query-time outcomes based on resolver-visible context and per-answer metadata, yielding different observable results for the same query under different conditions.

These semantics may be realized through a wide variety of mechanisms, configuration models, and representations, and are often described informally as \textit{traffic steering}. Despite their widespread deployment, the resolver-visible meaning of such behavior has not been formalized independently of particular implementations.

A key observation underlying this work is that authoritative DNS response selection is not an open-ended design space. At the authoritative interface, the DNS protocol imposes semantic constraints including finiteness of responses, RRset atomicity, totality of resolver-visible behavior, termination, cacheability over bounded time intervals, and restriction to resolver-visible query context and authoritative candidate data. Although each requirement is familiar in isolation, together they constrain different components of the semantic model and determine the admissible input domain, outcome space, and class of response-selection functions. The resulting function class is induced by DNS protocol semantics and is independent of how any particular realization implements, encodes, or exposes the behavior.

We formalize this protocol-induced class as DNS-admissible response-selection functions and prove that every such function admits a finite compositional normal form over admissible evaluation contexts and finite structured candidate domains, comprising individual answers and finite collections generated from them. The normal form first distinguishes among observationally relevant combinations of query context and candidate metadata, then reduces behavior within each case to restriction and selection over the corresponding candidate domain. The existence of this finite decomposition establishes expressive boundedness.

Expressive boundedness forms the bridge between protocol semantics and algebraic structure. The finite normal forms expose operations and composition laws that admit a natural semiring interpretation, enabling principled reasoning about equivalence, expressiveness, approximation, and semantic collapse. Concrete authoritative systems, configuration models, and serialized encodings---for example, provider-specific steering languages, intermediate representations, and DNS transfer representations---realize behaviors drawn from the same bounded semantic domain. They can therefore be modeled uniformly as semantic restrictions of that domain and analyzed within a common algebraic framework, differing in the behaviors they admit and the distinctions they preserve rather than constituting unrelated collections of features.

The argument proceeds as follows. \autoref{sec:constraints-and-resp-selection} derives the semantic constraints imposed by the DNS protocol and explains how they determine the admissible mathematical setting. \autoref{sec:formal-model} formalizes authoritative response selection within that setting and establishes expressive boundedness. \autoref{sec:concrete-realizations} defines the shared semantic domain and models concrete systems, configuration models, and representations as restrictions of that domain. \autoref{sec:algebraic-structure} exposes the domain's intrinsic compositional structure and provides its algebraic interpretation. \autoref{sec:realizations-as-algebraic-relations} develops exact representability, semantic collapse, approximation, and lowerability between realizations. Finally, \autoref{sec:implications-and-consequences} discusses the consequences of this perspective for equivalence, portability, semantic honesty, and future representations of authoritative DNS behavior.

\section{Related Work}
This work sits at the intersection of DNS protocol semantics, operational response-selection mechanisms, and algebraic models of compositional behavior. Rather than proposing a new response-selection mechanism, configuration language, or optimization, it makes explicit a semantic structure that is already implicitly enforced by the DNS protocol itself. As such, its relationship to prior work is primarily one of \emph{extraction and formalization} rather than extension or comparison.

\subsection{DNS Standards and Protocol Semantics}
The Domain Name System is specified through a long-standing series of RFCs, beginning with RFC~1034 and RFC~1035, and extended by later documents such as RFC~2181 and RFC~6891. These standards precisely define resolver-visible behavior, including message structure, RRset semantics, truncation, cacheability, and error handling.

While the DNS standards rigorously specify \emph{what} responses mean at the protocol boundary, they do not provide a system-independent semantic model of how authoritative servers select among multiple candidate answers at query time. Response selection logic is described operationally, in terms of message construction and resolver behavior, rather than as a semantic space amenable to equivalence, restriction, or approximation reasoning.

This work builds directly on the DNS standards by treating resolver-visible behavior as the sole semantic boundary, and by deriving constraints on admissible response-selection behavior directly from protocol requirements. The resulting model does not extend or reinterpret the standards; it formalizes semantic consequences that are already implicit in them.

\subsection{Operational and Provider-Specific Response Selection}
Modern authoritative DNS systems commonly support mechanisms described informally as ``traffic steering,'' including geographic routing, weighted selection, health-based filtering, priority ordering, and affinity-based behavior. These mechanisms are widely documented in provider-specific configuration languages and operational guides.

Such systems differ substantially in structure, terminology, and feature sets, and their behavior is typically reasoned about within the confines of a single provider model. While these mechanisms are operationally sophisticated, they are not generally presented in a form that supports system-independent semantic comparison, equivalence checking, or principled approximation across providers.

This work does not attempt to catalog or compare specific steering features. Instead, it abstracts over all such mechanisms by characterizing the space of resolver-visible behaviors permitted by DNS itself. Provider systems and configuration models are considered only insofar as they admit or restrict elements of this shared semantic domain.

\subsection{Formal Models and Algebraic Structure}
Several strands of prior work have applied algebraic and semantic techniques to networking, routing, and policy systems. In these settings, algebraic abstractions are typically introduced as modeling or control mechanisms, enabling reasoning about composition, reachability, or policy interaction.

Algebraic structures such as monoids and semirings are widely used in computer science to model compositional systems, including program semantics, weighted automata, and network control. Prior work in networking has applied similar abstractions to reason about routing behavior, policy composition, and centralized control in distributed systems.

The present work differs in both scope and motivation. Rather than proposing a language, control abstraction, or verification framework, it derives semantic structure directly from protocol-imposed constraints and treats implementations only as realizations of resolver-visible behavior. The algebraic structure identified here is not imposed a priori as a modeling choice; it arises as a consequence of the expressive bounds induced by DNS protocol semantics at the authoritative interface.

Similar algebraic abstractions have been used in other domains (e.g., routing and program analysis), but arise here for fundamentally different reasons: not from language or control-plane design, but as an intrinsic semantic consequence of DNS protocol constraints.

Recent work by Nevatia, Liu, and Basin \cite{Nevatia_2025} formalizes DNS semantics using algebraic abstractions in order to enable verification of DNS configurations and protocol behaviors. Their model treats DNS as a distributed protocol whose execution can be analyzed for reachability and safety properties. By contrast, the present work characterizes the space of resolver-visible authoritative response-selection semantics itself, focusing on expressive boundedness, semantic collapse, and approximation across realizations, independent of any particular execution or verification model.

\section{DNS Constraints and Authoritative Response Selection}
\label{sec:constraints-and-resp-selection}
This section derives semantic constraints on authoritative response selection directly from the DNS protocol. These constraints delimit the class of behaviors that any authoritative system may exhibit at the resolver-visible interface, independent of implementation strategy.

\subsection{Scope and Observation Boundary}
We restrict attention to authoritative DNS behavior at the protocol boundary. The semantics of interest are defined solely by the responses emitted by an authoritative system in reply to queries. Internal execution order, control-plane state, deployment topology, and operational heuristics are outside the model except insofar as they influence resolver-visible outcomes.

Accordingly, two authoritative systems are semantically equivalent if they emit indistinguishable DNS responses for all admissible queries. This restriction follows the DNS standards, which define semantics entirely in terms of on-the-wire behavior rather than internal computation.

\subsection{Protocol-Imposed Constraints}
Before formalizing authoritative response selection, we enumerate the semantic constraints imposed by the DNS protocol itself. These constraints delimit the class of behaviors that any authoritative system may exhibit at the resolver-visible interface.

For authoritative DNS, the protocol enforces the following constraints on query-time response selection:

\begin{itemize}
    \item \textbf{C1: Finiteness of candidate answers and responses.}
    Authoritative response selection operates over finite candidate sets and produces finite responses.

    \item \textbf{C2: RRset atomicity.}
    Observable semantics are defined over whole RRsets; partial or incremental RRset results are not meaningful.

    \item \textbf{C3: Totality of resolver-visible behavior.}
    For every admissible query, authoritative resolution yields a defined protocol outcome.

    \item \textbf{C4: Termination (well-founded evaluation).}
    Response selection must terminate; non-terminating evaluation is not permitted by DNS semantics.

    \item \textbf{C5: Time-bounded validity (cacheability).}
    Authoritative responses are intended to remain semantically valid over a bounded time interval.

    \item \textbf{C6: Dependence on resolver-visible inputs only.}
    Only resolver-visible query context and authoritative answer data may influence observable behavior.
\end{itemize}

These constraints act on different parts of the semantic model. Finiteness constrains the size of the candidate domain and outcome space. RRset atomicity constrains the structure of admissible outcomes. Totality constrains the definedness of the response-selection function. Termination constrains the admissible evaluation class. Cacheability constrains temporal dependence. Observability constrains the semantic input signature, not the operational mechanisms that populate it. Taken together, they determine not merely individual protocol requirements, but the admissible domain, codomain, and class of functions between them.

Importantly, these are not design choices of particular systems. They follow directly from DNS protocol semantics and apply uniformly to all authoritative implementations \cite{rfc1034,rfc1035,rfc2181,rfc6891}. We do not assume that they uniquely determine a single response-selection strategy, only that they bound the space of resolver-visible behaviors permitted by DNS.

The remainder of this section justifies each constraint with reference to the DNS standards and explains its semantic consequences.

\subsection{Finiteness of Candidate Answers and Responses}

\paragraph{Constraint.}
Authoritative response selection operates over finite candidate sets and produces finite responses.

\paragraph{RFC basis.}
RFC 1035 restricts DNS message size and defines truncation semantics:
\begin{quote}
``Messages carried by UDP are restricted to 512 bytes\ldots Longer messages are truncated and the TC bit is set.'' \cite[\S4.2.1]{rfc1035}
\end{quote}

Additionally, resource records and RRsets are finite by construction \cite[\S3.2.1]{rfc1035}.

\paragraph{Interpretation.}
The relevant finiteness constraint is semantic rather than merely transport-level. DNS over TCP does not turn DNS into an open-ended stream of resource records; it carries length-prefixed DNS messages, and transfer mechanisms such as AXFR consist of discrete DNS messages rather than an unbounded RR-level semantic stream. Thus, TCP may change transport capacity, but it does not enlarge the resolver-visible semantic domain.

Because DNS responses are finite protocol objects and RR encoding is finite, any authoritative response must consist of a finite DNS-admissible outcome. Consequently, at evaluation time, response selection operates over a finite candidate domain and emits a finite resolver-visible result.

\paragraph{Semantic consequence.}

This justifies modeling authoritative response selection as operating, at evaluation time, over a finite set $A$ of primitive candidate answers together with finite structured collections generated from $A$.

Importantly, the DNS protocol requires finiteness of candidate data and resolver-visible outcomes, but it does not require semantic operators to act only on singleton answers. Finite collections of answers, such as groups, pools, or region-associated sets, may also serve as semantic objects so long as observable outcomes remain finite and DNS-admissible.

Accordingly, authoritative response selection is more precisely understood as operating over a finite structured candidate domain generated from $A$, rather than over isolated answer instances alone.

\subsection{Atomicity of RRsets}

\paragraph{Constraint.}
Observable DNS semantics are defined over whole RRsets, not partial or incremental results.

\paragraph{RFC basis.}
RFC 2181 specifies RRset atomicity:
\begin{quote}
``The TC bit should only be set when an RRSet\ldots could not be included in its entirety.'' \cite[\S5.2]{rfc2181}
\end{quote}

\paragraph{Interpretation.}
DNS does not permit partial RRset semantics: an RRset is either present in full or absent. Any truncation or omission is observable only at the RRset level.

\paragraph{Semantic consequence.}
RRset atomicity constrains the shape of admissible outcomes. Authoritative response selection can be modeled as choosing DNS-admissible outcomes over a finite structured candidate domain, including subsets, ordered collections, or distributions over such collections, but not arbitrary streams, prefixes, or partial RRset fragments. Truncation does not create a semantic notion of partial delivery; it signals that the requested RRset could not be carried in full.

Thus, RRset atomicity constrains the resolver-visible output boundary while still allowing internal evaluation over finite structured candidate objects such as groups, pools, or region-associated collections.

\subsection{Totality of Resolver-Visible Behavior}

\paragraph{Constraint.}
Authoritative response selection is total over admissible queries.

\paragraph{RFC basis.}
RFC 1034 defines the outcome of resolution attempts:
\begin{quote}
``The resolver will have one of the following: an answer, a referral, or an error.'' \cite[\S4.3.1]{rfc1034}
\end{quote}

\paragraph{Interpretation.}
From the resolver’s perspective, authoritative resolution always yields a defined protocol outcome. There is no notion of an undefined or non-terminating response at the semantic level.

\paragraph{Semantic consequence.}
Response selection may therefore be modeled as a total function over admissible inputs. When restricted to selection among candidate answers, this function always produces a (possibly empty) admissible response.

\subsection{Termination and Well-Founded Evaluation}

\paragraph{Constraint.}
Authoritative response selection must terminate.

\paragraph{RFC basis.}
RFC 1034 explicitly treats resolution loops as errors:
\begin{quote}
``If a domain name server is so misconfigured as to contain a loop, it can be signalled as an error.'' \cite[\S5.2.2]{rfc1034}
\end{quote}

\paragraph{Interpretation.}
The DNS protocol disallows non-terminating resolution behavior. Any query-time decision process that fails to terminate violates protocol semantics.

\paragraph{Semantic consequence.}
Response selection semantics may be modeled as finite, well-founded evaluation—excluding recursive or unbounded control flow from the semantic model.

\subsection{Cacheability and Time-Bounded Validity}

\paragraph{Constraint.}
Authoritative responses are intended to be stable over bounded time intervals.

\paragraph{RFC basis.}
RFC 1035 defines TTL semantics:
\begin{quote}
``TTL specifies the time interval that the resource record may be cached.'' \cite[\S3.2.1]{rfc1035}
\end{quote}

RFC 2181 reinforces cacheability requirements:
\begin{quote}
``Data from authoritative servers should be cacheable.'' \cite[\S5.2]{rfc2181}
\end{quote}

\paragraph{Interpretation.}
While authoritative systems may vary responses over time, DNS explicitly defines a validity interval over which responses are intended to be reused by resolvers. This precludes semantics that depend on unconstrained, per-query mutable state invisible to resolvers.

This does not imply that authoritative implementations are stateless, nor that they do not correlate responses across queries. Many systems employ mechanisms such as session affinity, consistent hashing, or so-called ``sticky'' selection strategies, in which responses are correlated based on aspects of the query origin (e.g., resolver IP address or network prefix).

Crucially, such mechanisms do not introduce additional semantic degrees of freedom beyond those already permitted by the DNS protocol. Any correlation that is stable and observable to resolvers must be keyed on resolver-visible query attributes and must respect cacheability constraints. Conversely, correlations based on unconstrained, hidden mutable state cannot be relied upon by resolvers and therefore do not constitute distinct DNS semantics.

\paragraph{Semantic consequence.}
Response selection may be treated as a function of resolver-visible query context and available answer metadata, without reference to hidden, unbounded history.

Importantly, this does \emph{not} assume strict determinism—only bounded semantic variability consistent with DNS caching semantics.

\subsection{Observable Inputs Only}

\paragraph{Constraint.}
Only resolver-visible query context and authoritative answer data may influence response selection semantics.

\paragraph{RFC basis.}
DNS semantics are defined entirely in terms of message contents and authoritative data; unknown EDNS options must be ignored \cite[\S6.1.2]{rfc6891}, and unrecognized metadata has no protocol-defined meaning.

\paragraph{Interpretation.}
Any internal state or computation that does not manifest in resolver-visible responses is semantically irrelevant under DNS.

\paragraph{Semantic consequence.}
Authoritative response selection semantics can be modeled exclusively in terms of observable inputs and outputs.

\section{Formal Model and Expressive Boundedness}
\label{sec:formal-model}
Having derived the protocol-imposed constraints on authoritative response selection in \autoref{sec:constraints-and-resp-selection}, we now formalize the class of behaviors admissible under those constraints. The objective of this section is twofold: first, to define a semantic model that captures exactly the resolver-visible behavior permitted by DNS; and second, to show that this class of behaviors is expressively bounded.

\subsection{Semantic Domains}
We model authoritative response selection at the resolver-visible interface.

Let $Q$ denote the set of admissible resolver-visible query contexts. By *query context* we mean the information carried with the DNS query message itself and any protocol-defined extensions that are visible to the authoritative server at query time. This includes, at a minimum, the queried name and type, and may include other message fields or extension data defined by the DNS protocol.

We explicitly exclude information not conveyed by the query message, such as internal server state, wall-clock time, or implementation-specific metadata, except insofar as such information is reflected in resolver-visible query attributes. We do not distinguish between statically enumerated and dynamically constructed RRsets: any construction process that yields a finite, cacheable RRset keyed on resolver-visible inputs induces a DNS-admissible function and is therefore already included in $\mathcal{D}$.

Let $A$ denote a finite set of primitive candidate answers available at evaluation time. Finiteness of $A$ follows directly from Constraint C1.

Authoritative response selection need not operate only on individual elements of $A$. DNS-admissible semantics may also operate over finite structured collections generated from $A$, such as grouped answer sets, pools, region-associated collections, or other finite compositions of candidate answers. Singleton answers are treated as degenerate collections.

Let $\mathcal{C}(A)$ denote the finite structured candidate domain generated from $A$. Elements of $\mathcal{C}(A)$ may include singleton answers, finite subsets, ordered collections, grouped collections, or other finite structures whose observable outputs remain DNS-admissible.

Let $M$ denote metadata associated with primitive answers and, where applicable, with structured candidate objects in $\mathcal{C}(A)$. Such metadata may include attributes or annotations that influence restriction or selection.

Let $\Delta(\mathcal{C}(A))$ denote the set of admissible resolver-visible outcomes over the structured candidate domain. $\Delta(\mathcal{C}(A))$ is finite for fixed $A$ and fixed finite structure. The specific choice of outcome representation is not material to our results. Any finite, RRset-atomic (Constraint C2) outcome domain over $\mathcal{C}(A)$ suffices, including subsets, multisets, permutations, or finite-support distributions, provided that emitted DNS records are selected from existing RRset members.

\subsection{Materialization Boundary and Opaque Inputs}
\label{sec:materialization-boundary}

The definition of $A$, $\mathcal{C}(A)$, and $M$ establishes a materialization boundary. Mechanisms such as health checks, data feeds, pool expansion, ALIAS flattening, upstream lookup, or provider-specific answer synthesis may determine which candidate answers are available or which metadata values are attached to them. But once those mechanisms have produced the evaluation environment, response selection still has the form:
$$
F : (Q, \mathcal{C}(A), M) \rightarrow \Delta(\mathcal{C}(A)).
$$

Thus, external or implementation-specific mechanisms may change $A$, $\mathcal{C}(A)$, or $M$, but they do not change the semantic shape of $F$. They materialize the finite candidate domain or annotate it; they do not introduce a new class of DNS response-selection semantics.

This distinction is important for opaque mechanisms such as external data feeds. If a feed’s internal logic is not represented, the model does not attempt to infer or preserve that internal computation. It preserves only the resolver-visible dependency on the feed’s materialized output. In such cases, partitioning is over the supplied values in $(Q,M)$, not over the hidden logic that produced them.

Consequently, opaque inputs preserve expressive boundedness but may reduce portability and explainability. A behavior depending on a feed-derived metadata field is exactly portable to another realization only if that realization, or an associated control-plane process, supplies equivalent materialized metadata with compatible stability and cacheability assumptions. Otherwise, translation must be classified as approximate or non-representable.

In short: feeds and health checks may change the candidate domain or metadata, but they do not escape the bounded DNS-admissible semantic domain.

\subsection{DNS-Admissible Response-Selection Functions}
We now define the class of response-selection behaviors permitted by DNS.

\begin{definition}[DNS-Admissible Response Selection]
\label{def:dns-admissible-resp-selection}
A DNS-admissible response-selection function is a total function
$$
F : (Q, \mathcal{C}(A), M) \rightarrow \Delta(\mathcal{C}(A))
$$
satisfying the following conditions:
\begin{enumerate}
\item \textbf{Totality:} $F$ is defined for all admissible inputs (Constraint C3).
\item \textbf{Finiteness:} $F : (Q, \mathcal{C}(A), M)$ yields a finite outcome for all inputs (Constraint C1).
\item \textbf{RRset atomicity:} Outcomes correspond to whole RRsets (Constraint C2).
\item \textbf{Termination:} Evaluation of $F$ is finite and well-founded (Constraint C4).
\item \textbf{Cache-consistency:} Variability of $F$ is consistent with bounded cache validity (Constraint C5).
\item \textbf{Observability:} $F$ depends only on resolver-visible query context, authoritative answer data, and finite structured candidate objects derived from that data, represented as $(Q,\mathcal{C}(A),M)$ (Constraint C6).
\end{enumerate}
\end{definition}

This definition abstracts away all implementation detail, capturing only resolver-visible semantics. The DNS protocol defines a set of admissible resolver-visible behaviors; each such behavior induces a function of the form above.

\subsection{Structural Consequences of the Constraints}
The constraints in \autoref{sec:constraints-and-resp-selection} jointly impose strong structural restrictions on DNS-admissible response-selection functions.

First, because candidate sets and outcomes are finite, and evaluation must terminate, response selection cannot express unbounded computation or recursion. Second, because semantics are defined only in terms of resolver-visible inputs and outputs, distinctions that do not manifest in observable responses are semantically irrelevant. Finally, cacheability bounds the degree to which response selection may vary over time or across queries without violating DNS semantics.

Together, these properties imply that authoritative response selection does not constitute an open-ended control system. 
Instead, it inhabits a closed semantic space determined by finite conditional discrimination over observable inputs followed by restriction and selection over a finite structured candidate domain.

\subsection{Expressive Boundedness}
We now state the central boundedness result.

\begin{theorem}[Expressive Boundedness]
\label{thm:expressive-boundedness}
The class of DNS-admissible response-selection functions is expressively bounded: every such function can be represented, up to semantic equivalence, as a finite composition of conditional restrictions over $(Q, M)$ together with selection over finite structured candidate collections generated from the finite candidate-answer set $A$.
\end{theorem}

\paragraph{Informal justification.}
Because only finitely many observationally distinguishable cases over $Q$ and $M$ may influence resolver-visible outcomes (by Constraints C1, C4, and C6), response selection may be viewed as partitioning the input space into a finite set of regions. Within each region, behavior reduces to restriction and selection over the finite structured candidate domain $\mathcal{C}(A)$, subject to RRset atomicity at the resolver-visible output boundary. The absence of unbounded state and the requirement of termination ensure that this decomposition is finite and well-founded.

The compositional decomposition guaranteed by \autoref{thm:expressive-boundedness} is not necessarily unique: distinct compositions may denote the same DNS-admissible response-selection function.

A formal proof is deferred to Appendix A.

\subsection{Implications}
\autoref{thm:expressive-boundedness} establishes a fundamental semantic limitation on authoritative DNS response selection. Apparent diversity among concrete systems arises not from unbounded semantics, but from different ways of realizing restricted subsets of this bounded space.

In the next section, we make this structure explicit by introducing a small compositional basis sufficient to represent all DNS-admissible response-selection functions.

The boundedness result of this section is purely semantic: it characterizes the space of resolver-visible behaviors permitted by DNS, independent of representation. Subsequent sections exploit this boundedness to introduce algebraic structure and reason about approximation and expressiveness, but those results rely critically on the functional characterization established here.

\section{Concrete Realizations of DNS-Admissible Semantics}
\label{sec:concrete-realizations}
\autoref{sec:formal-model} established that authoritative DNS response selection is expressively bounded: every DNS-admissible resolver-visible behavior admits a finite normal form consisting of conditional restriction followed by selection among a finite candidate set. This result characterizes the full semantic envelope permitted by the DNS protocol, independent of any particular implementation or configuration language.

In practice, however, authoritative DNS behavior is realized only through specific systems, standards, or deployment contexts, each of which admits only a restricted subset of this envelope. This section introduces a semantic notion of \textbf{concrete realization} that captures this restriction explicitly. By modeling realizations as semantic subsets rather than syntactic artifacts, we decouple questions of expressiveness, compatibility, and correctness from any particular configuration language or implementation strategy.

\subsection{The Semantic Domain of DNS Steering}

Let
$$
\mathcal{D} \;=\; \{\, F \mid F : (Q, \mathcal{C}(A), M) \rightarrow \Delta(\mathcal{C}(A)),\; F \text{ is DNS-admissible} \,\}
$$
denote the semantic domain of all DNS-admissible response-selection functions as defined in \autoref{sec:formal-model}.

By \autoref{thm:expressive-boundedness}, every element of $\mathcal{D}$ admits a finite normal form consisting of conditional restriction over $(Q, M)$ followed by selection over a finite structured candidate domain $\mathcal{C}(A)$. No additional expressive forms are permitted by the DNS protocol.

This domain is semantic rather than syntactic: it classifies resolver-visible behaviors, not configurations, algorithms, or evaluation strategies.

This formulation also explains provider grouping constructs. Groups, pools, and regions are not treated as additional primitive operators. They are finite structured candidate objects in $\mathcal{C}(A)$ over which the existing operation families—restriction and selection—may act. Thus, introducing structured candidate domains generalizes the semantic objects of the algebra without expanding the primitive operator set.

\subsection{Concrete Realizations as Semantic Restrictions}

A \textbf{concrete realization} is defined as a restriction of the full semantic domain.

\begin{definition}[Concrete Realization]
A concrete realization $\mathcal{R}$ is a subset
$$
\mathcal{D}_{\mathcal{R}} \subseteq \mathcal{D}
$$
together with a sound semantic interpretation that maps configuration constructs, rules, or policies into elements of $\mathcal{D}_{\mathcal{R}}$.
\end{definition}

Intuitively, a concrete realization specifies:
- which DNS-admissible behaviors may be expressed, and
- which semantic distinctions may be observed, enforced, or preserved.

Crucially, a concrete realization does \textbf{not} introduce new semantic functions. All resolver-visible mappings it admits are elements of $\mathcal{D}$; the realization merely restricts which elements of $\mathcal{D}$ are accessible.

A concrete realization need not itself perform authoritative query-time evaluation; serialized representations, intermediate formats, or zone-file encodings that constrain which elements of $\mathcal{D}$ are admissible also constitute realizations in this sense.

\subsection{Exact Realizability}

Given a target semantic behavior $f \in \mathcal{D}$ — that is, a specified resolver-visible mapping $(Q,A,M)\to\Delta(A)$ — and a concrete realization $\mathcal{R}$, the most basic question is whether that behavior can be realized \textit{without semantic loss}.

\begin{definition}[Exact Realizability]
A DNS-admissible response-selection function
$$
f \in \mathcal{D}
$$
is \textbf{exactly realizable} under a concrete realization $\mathcal{R}$ if
$$
f \in \mathcal{D}_{\mathcal{R}}.
$$
\end{definition}

Exact realizability is a property of the resolver-visible behavior: it depends only on the meaning of the target behavior $f$ and the semantic restrictions imposed by $\mathcal{R}$. It is independent of any particular lowering, compilation, or synthesis procedure.

\subsection{Consequences of Expressive Boundedness}

The boundedness result of \autoref{sec:formal-model} imposes strong structure on the realizability problem. Because every element of $\mathcal{D}$ admits a finite conditional–selection normal form:
- realizability questions are finite and well-defined,
- semantic comparison is tractable,
- and realizations may be analyzed compositionally.

Moreover, because concrete realizations are defined as semantic restrictions rather than syntactic languages, incompatibilities between systems can be understood as genuine semantic gaps rather than artifacts of representation.

These properties form the foundation for reasoning about equivalence, collapse, and approximation between realizations, which we develop in subsequent sections.

\subsection{Discussion}

By modeling authoritative DNS systems, standards, and deployment contexts uniformly as concrete realizations of a shared semantic domain, we separate questions of \textbf{what DNS permits} from questions of \textbf{what particular systems admit}. This separation clarifies the sources of incompatibility between systems and enables principled reasoning about portability and correctness without appeal to feature parity or ad hoc translation.

In the next section, we expose additional structure on $\mathcal{D}$ and its realizations, enabling algebraic reasoning about expressiveness, restriction, and composition.

\section{Algebraic Structure of Concrete Realizations}
\label{sec:algebraic-structure}

\autoref{sec:concrete-realizations} modeled authoritative DNS systems, configuration models, and other semantic encodings uniformly as concrete realizations of DNS-admissible behavior—semantic restrictions of a shared domain $\mathcal{D}$ of resolver-visible response-selection functions. This section exposes additional structure on $\mathcal{D}$ and its realizations, enabling algebraic reasoning about expressiveness, restriction, and composition.

The key observation is that the normal forms guaranteed by \autoref{thm:expressive-boundedness} are not merely finite, but \textbf{compositional}. Conditional restriction and selection compose in a disciplined way: they are closed under composition, associative, and interact distributively. These properties induce an algebraic structure on $\mathcal{D}$, which we make explicit below. Concrete realizations inherit this structure as restricted subspaces, allowing expressiveness and compatibility to be analyzed algebraically rather than syntactically.

\subsection{Compositional Operations on $\mathcal{D}$}

By \autoref{thm:expressive-boundedness}, every function $F \in \mathcal{D}$ admits a finite representation as a composition of conditional restriction and selection. These two operation families induce natural composition operators on $\mathcal{D}$:
\begin{itemize}
\item \textbf{Conditional restriction}, which limits the applicability of a behavior based on predicates over $(Q, M)$;
\item \textbf{Selection composition}, which combines alternative admissible outcomes over a shared candidate set.
\end{itemize}

These operations preserve DNS admissibility: composing DNS-admissible behaviors using conditional restriction and selection yields another DNS-admissible behavior. Consequently, $\mathcal{D}$ is closed under these compositions.

\subsection{Algebraic Interpretation}

The closure and distributive interaction of conditional restriction and selection induce an algebraic structure on $\mathcal{D}$. In particular:
- selection behaves as an additive operation, combining alternative admissible outcomes;
- conditional restriction behaves as a multiplicative operation, gating behaviors by predicates;
- both operations are associative and admit identity elements corresponding to unconditional pass-through and empty behavior.

These properties follow directly from the extensional definition of DNS-admissible functions and the normal forms established in \autoref{sec:formal-model}. Intuitively, this follows because conditional restriction corresponds to finite predicate gating and selection corresponds to disjoint choice over finite outcomes. Formal verification of these algebraic properties is straightforward and can be given by structural induction on the normal forms of \autoref{thm:expressive-boundedness}.

Together, these properties endow $\mathcal{D}$ with the structure of a \textbf{semiring}—a minimal algebra capturing composition without inverses. We introduce this identification to make precise which forms of comparison, restriction, and approximation are semantically well-founded; the subsequent sections exploit these consequences. A precise algebraic formulation is given in Appendix B; here we focus on its semantic implications.

Importantly, this structure is \textbf{not imposed} by representation choices. It arises directly from the compositional normal forms required by DNS semantics. The algebraic perspective therefore reflects intrinsic structure of the DNS-admissible semantic space rather than an artifact of a particular intermediate representation.

\subsection{Concrete Realizations as Subalgebras}

Each concrete realization $\mathcal{R}$, defined in \autoref{sec:concrete-realizations} as a subset $\mathcal{D}_{\mathcal{R}} \subseteq \mathcal{D}$, inherits the algebraic structure of $\mathcal{D}$ subject to its restrictions.

In general, $\mathcal{D}_{\mathcal{R}}$ is closed under some but not all compositional operations. For example, a realization may:
- restrict which predicates may be used for conditional restriction,
- limit the available selection strategies,
- or impose fixed evaluation orders that collapse otherwise distinct compositions.

Such restrictions induce algebraic restrictions of $\mathcal{D}$, often—but not exclusively—subalgebras. Expressiveness differences between systems therefore arise from differences in the algebraic restrictions they impose, whether by exclusion of elements, identification of elements, or both.

This framing makes explicit that no concrete realization exceeds the expressive power of DNS itself; realizations differ only by omission.

\subsection{Expressiveness and Semantic Restriction}

The algebraic view provides a principled notion of expressiveness.

A realization $\mathcal{R}_1$ is at least as expressive as $\mathcal{R}_2$ if $\mathcal{D}_{\mathcal{R}_2}$ embeds into $\mathcal{D}_{\mathcal{R}_1}$ as a subalgebra. Strict expressiveness loss corresponds to the absence of closure under certain compositions.

From this perspective:
\begin{itemize}
\item apparent feature gaps correspond to missing generators or closure properties,
\item “unsupported” configurations correspond to elements of $\mathcal{D}$ lying outside a realization’s subalgebra,
\item and semantic collapse arises when distinct elements of $\mathcal{D}$ are identified under a restricted algebra.
\end{itemize}

This reframes expressiveness analysis as an algebraic comparison problem rather than a feature-by-feature enumeration.

\subsection{Implications}

By exposing the algebraic structure underlying DNS-admissible response selection, we obtain a uniform framework for reasoning about concrete systems:
\begin{itemize}
\item \textbf{Equivalence} becomes equality within a shared algebra.
\item \textbf{Restriction} becomes passage to a subalgebra.
\item \textbf{Loss of expressiveness} becomes algebraic collapse.
\item \textbf{Compatibility and translation} become questions of homomorphism between algebras.
\end{itemize}

These observations provide the foundation for reasoning about approximation, lowerability, and semantic preservation, which we develop in the next section.

\subsection{Discussion}

The algebraic interpretation of DNS-admissible response-selection semantics does not replace the functional characterization of \autoref{sec:formal-model}; it depends on it. The expressive boundedness result guarantees that the algebra spans the entire semantic domain, while the algebraic structure enables reasoning that would be cumbersome or opaque at the level of raw functions.

In this sense, the algebra is \textbf{discovered}, not chosen. It reflects the compositional structure enforced by DNS protocol semantics and made explicit by the expressive boundedness theorem.

\section{Realizations as Algebraic Restrictions}
\label{sec:realizations-as-algebraic-relations}

Sections~\ref{sec:formal-model}--\ref{sec:algebraic-structure} established a bounded semantic domain $\mathcal{D}$ of DNS-admissible response-selection semantics and exposed its intrinsic algebraic structure. We now use that structure to reason precisely about how such semantics relate across concrete realizations, and why translation, preservation, and approximation are fundamentally constrained.

\subsection{Realizations as Restricted Semantic Domains}

Recall from \autoref{sec:concrete-realizations} that a concrete realization $\mathcal{R}$ admits only a subset of DNS-admissible semantics:
$$
\mathcal{D}_{\mathcal{R}} \subseteq \mathcal{D}.
$$
This restriction is semantic rather than syntactic: it specifies which resolver-visible behaviors may be expressed or preserved, independent of how those behaviors are represented or evaluated.

Algebraically, such restriction cannot in general be modeled by substructure alone. Concrete realizations restrict $\mathcal{D}$ in two independent ways:
\begin{enumerate}
\item \textbf{Forbidden expressions (substructure).}
    Certain DNS-admissible semantics may be entirely unexpressible within a realization. This arises when the realization forbids particular predicates over $(Q,M)$, disallows forms of selection over $(\mathcal{C}(A),M)$, or imposes invariants that exclude entire classes of behavior. These effects correspond to removing elements from $\mathcal{D}$.

\item \textbf{Identified expressions (congruence).}
    Distinct elements of $\mathcal{D}$ may become observationally indistinguishable under a realization. When a realization fails to preserve certain distinctions—such as ordering, weighting precision, or conditional discrimination—multiple DNS-admissible semantics collapse to the same observable behavior.
\end{enumerate}
Substructure alone captures loss of expressibility but cannot represent identification of remaining elements. Quotienting alone captures identification but cannot forbid expressions outright. Concrete realizations may impose either or both effects, and therefore correspond to algebraic restrictions involving both substructure and congruence.

\subsection{Semantic Collapse and Irreversibility}

When a realization identifies distinct elements of $\mathcal{D}$, semantic collapse occurs. Two DNS-admissible response-selection semantics that are distinct in $\mathcal{D}$ may be observationally indistinguishable within $\mathcal{R}$.

Such collapse is irreversible. Once distinctions are eliminated—whether by restricting predicates, collapsing candidate sets, or weakening selection guarantees—no further composition within $\mathcal{D}_{\mathcal{R}}$ can recover them. This irreversibility is a direct consequence of DNS semantics and is reflected algebraically by the absence of cancellation or inverses.

Algebraically, semantic collapse may be understood as identification under a realization-specific congruence on $\mathcal{D}$. We do not require an explicit quotient construction; it suffices to observe that collapse induces many-to-one mappings from $\mathcal{D}$ into the semantics admitted by the realization.  

\subsection{Exact Representability}

Given a target DNS-admissible semantic $f \in \mathcal{D}$ and a realization $\mathcal{R}$, the most basic question is whether $f$ can be realized without semantic loss.

\begin{definition}[Exact Representability]
A DNS-admissible response-selection semantic $f \in \mathcal{D}$ is \textbf{exactly representable} under a realization $\mathcal{R}$ if f survives the realization’s restrictions without collapse—that is, if f lies within $\mathcal{D}_{\mathcal{R}}$ and is not identified with any distinct semantic under the realization’s induced congruence.
\end{definition}

Exact representability is a semantic property. It depends only on the meaning of $f$ and the algebraic restrictions imposed by $\mathcal{R}$, not on any particular lowering, compilation, or synthesis procedure.

Failure of exact representability reflects a genuine expressive mismatch between the target semantic and the realization.

\subsection{Approximation and Homomorphisms}

When exact representability fails, approximation may still be possible.

Algebraically, admissible approximation corresponds to the existence of a (possibly many-to-one) semiring homomorphism from $\mathcal{D}$ into the semantics admitted by $\mathcal{D}_\mathcal{R}$. Such homomorphisms reflect semantic collapse and must be monotone with respect to restriction and identification: they may eliminate distinctions but cannot introduce new ones.

\subsection{Approximation Preorder}

Approximation is not merely existential; multiple admissible approximations may exist for a given target semantic. These approximations may be compared by the degree of semantic collapse they induce.

\begin{definition}[Approximation Preorder]
Fix a target semantic $f \in \mathcal{D}$ and a concrete realization $\mathcal{R}$.

Let $\mathrm{Approx}_{\mathcal{R}}(f) \subseteq \mathcal{D}_{\mathcal{R}}$ denote the set of semantics that approximate $f$ under $\mathcal{R}$.

Define a binary relation $\preceq_f$ on $\mathrm{Approx}_{\mathcal{R}}(f)$ by:
$$
g_1 \preceq_f g_2 \quad\text{iff}\quad g_1 \text{ preserves all resolver-visible distinctions of } f \text{ that } g_2 \text{ preserves}.
$$

Equivalently, $g_1 \preceq_f g_2$ if $g_1$ induces no greater semantic collapse of $f$ than $g_2$.
\end{definition}

The relation $\preceq_f$ is a \textbf{preorder}: it is reflexive and transitive but need not be antisymmetric. Minimal elements under $\preceq_f$ correspond to \textit{least-deviating} admissible approximations when such elements exist. Neither uniqueness nor existence of minimal approximations is guaranteed.

Intuitively, semantic collapse is quantified by the resolver-visible distinctions that remain observable. An approximation exhibits \textbf{\textit{less} collapse} than another if it \textbf{preserves \textit{more} of the distinctions} present in the target behavior—such as distinctions between query contexts, candidate answers, or selection outcomes. Conversely, an approximation exhibits \textit{more} collapse if it identifies cases that were previously distinguishable.

In this sense, approximation is ordered by \textbf{distinction preservation}, not by numeric distance. One approximation is strictly better than another when it preserves all the distinctions that the other preserves, and possibly more. When two approximations preserve different, incomparable sets of distinctions, neither is strictly better than the other.

\subsection{Lowerability}

Exact and approximate representability together determine \textbf{lowerability}.

\begin{definition}[Lowerability]
A DNS-admissible response-selection semantic $f \in \mathcal{D}$ is \textit{lowerable} to a realization $\mathcal{R}$ if there exists a semiring homomorphism $h: \mathcal{D} \to \mathcal{D}_\mathcal{R}$ such that $h(f) \in \text{Approx}_\mathcal{R}(f)$. 
\end{definition}

Lowerability is therefore a semantic predicate. It can, in principle, be determined independently of any concrete lowering algorithm. When lowerability fails, no correct implementation exists within the realization’s semantic constraints.

\section{Implications and Consequences}
\label{sec:implications-and-consequences}

The preceding sections establish that authoritative DNS response selection inhabits a bounded semantic domain $\mathcal{D}$, admits a disciplined algebraic structure, and is realized in practice only through restricted semantic projections. Making these facts explicit has several important consequences for reasoning about correctness, expressiveness, and portability.

\subsection{Boundedness as a Semantic Invariant}

The expressive boundedness result of \autoref{sec:formal-model} implies that authoritative DNS response selection is not an open-ended control system. All admissible behaviors factor through finite conditional discrimination followed by restriction and selection over finite structured candidate domains. No realization—regardless of complexity or implementation strategy—can exceed this envelope without violating DNS protocol semantics.

This observation shifts the focus of analysis away from feature enumeration and toward semantic restriction. Differences between systems arise not from fundamentally different semantic capabilities, but from how each realization restricts, collapses, or preserves elements of the same bounded domain.

\subsection{Equivalence and Semantic Collapse}

Because realizations may identify distinct DNS-admissible semantics, semantic equivalence is realization-dependent. Two behaviors that are distinct in $\mathcal{D}$ may be observationally indistinguishable under a particular realization, while remaining distinguishable under another.

The algebraic framing clarifies that such collapse is irreversible: once a distinction is eliminated by restriction or identification, no subsequent composition can recover it. This irreversibility is not an artifact of representation, but a direct consequence of DNS semantics and the absence of cancellation in the induced algebra.

As a result, equivalence checking must be performed relative to a realization. Global equivalence in $\mathcal{D}$ is strictly finer than equivalence under any concrete system.

\subsection{Lowerability and Honest Failure}

Lowerability analysis, as defined in \autoref{sec:realizations-as-algebraic-relations}, distinguishes genuine semantic incompatibility from implementation limitation. When no admissible homomorphic image of a target semantic exists within a realization, failure of lowering reflects a true expressive mismatch rather than an inadequacy of a particular translation strategy.

This framing makes failure informative. Rather than silently approximating or implicitly strengthening semantics, systems may report precisely which distinctions cannot be preserved and why. Such explanations are grounded in protocol-imposed constraints rather than provider-specific behavior.

\subsection{Approximation and Least Deviation}

When exact representability fails but approximation is possible, the approximation preorder of \autoref{sec:realizations-as-algebraic-relations} provides a principled basis for comparison. Approximations may be ordered by the degree of semantic collapse they induce relative to a target behavior, allowing least-deviating approximations to be identified when minimal elements exist.

Importantly, neither existence nor uniqueness of such minimal approximations is guaranteed. Incomparable approximations may reflect fundamentally different ways of weakening behavior—collapsing different distinctions that cannot be preserved simultaneously under the realization’s constraints.

This perspective avoids treating approximation as heuristic or ad hoc. Instead, it is governed by monotonicity and irreversibility inherent in DNS semantics.

\subsection{Separation of Semantics from Representation}

By grounding all reasoning in resolver-visible semantics, the framework cleanly separates questions of meaning from questions of syntax or implementation. Configuration languages, intermediate representations, and serialized encodings all participate as concrete realizations only insofar as they admit or restrict elements of $\mathcal{D}$.

This separation avoids conflating expressiveness with feature parity and prevents accidental overcommitment to semantics not enforced by structure. It also permits future representations or standards to be evaluated within the same semantic framework, independent of their surface form.

This separation also clarifies the role of black-box inputs. Opaque feeds, health systems, and synthesis mechanisms may be necessary to construct the evaluation environment, but their hidden implementation is not itself part of response-selection semantics unless it is represented as observable candidate data or metadata. Such mechanisms therefore do not expand $\mathcal{D}$; they define assumptions under which an element of $\mathcal{D}$ is realized. When those assumptions cannot be preserved across systems, the issue is not failure of boundedness, but loss of portability or explainability.

\subsection{Broader Implications}

Although this work focuses on authoritative DNS response selection, the methodological approach is more general. Protocol-imposed semantic constraints, once made explicit, often induce bounded expressive domains with tractable algebraic structure. Making these bounds explicit enables principled reasoning about equivalence, portability, and approximation across heterogeneous systems.

In the DNS case, this analysis clarifies long-standing ambiguity about the role and limits of so-called “traffic steering.” Rather than treating steering as an informal collection of techniques, it emerges as a well-defined semantic class shaped directly by protocol semantics.

\section{Conclusion}
\label{sec:conclusion}

This work shows that authoritative DNS response selection is not an open-ended design space, but a bounded semantic domain determined directly by protocol constraints. By making these bounds explicit, we show that all admissible resolver-visible behaviors factor through a finite conditional–selection structure over finite structured candidate domains and admit intrinsic algebraic organization.

Modeling systems and representations as semantic restrictions of this domain separates questions of meaning from questions of implementation, and explains expressiveness loss, equivalence, and approximation as consequences of restriction and collapse rather than feature disparity.

The result is not a new steering mechanism, but a clarified semantic foundation: what DNS permits, what systems restrict, and what can—and cannot—be preserved when moving between them.

\appendix
\section{Proof of Expressive Boundedness (\autoref{thm:expressive-boundedness})}
\label{app:expressive-boundedness-proof}
  
This appendix provides a formal proof of \autoref{thm:expressive-boundedness}. The proof relies solely on the protocol-imposed constraints derived in \autoref{sec:formal-model} and the definitions introduced in \autoref{sec:concrete-realizations}.

\subsection{Restatement of the Theorem}

\begin{theorem}[Expressive Boundedness]
Every DNS-admissible response-selection function
$$
F : (Q, \mathcal{C}(A), M) \rightarrow \Delta(\mathcal{C}(A))
$$

can be represented, up to semantic equivalence, as a finite composition of conditional restrictions over $(Q, M)$ together with restriction and selection over a finite structured candidate domain $\mathcal{C}(A)$ generated from $A$.
\end{theorem}

\subsection{Assumptions and Constraints}
\label{appsec:assumptions-and-constraints}

We assume the following properties of DNS-admissible response-selection functions, as established in \autoref{sec:formal-model}:
\begin{itemize}
\item \textbf{C1 (Finiteness):} Candidate sets and outcomes are finite.
\item \textbf{C2 (RRset atomicity):} Outcomes correspond to whole RRsets.
\item \textbf{C3 (Totality):} F is total over admissible inputs.
\item \textbf{C4 (Termination):} Evaluation of F is finite and well-founded.
\item \textbf{C5 (Time-bounded validity):} Variability is consistent with bounded cache validity.
\item \textbf{C6 (Observability):} $F$ depends only on resolver-visible query context, authoritative answer data, and finite structured candidate objects derived from that data, represented as $(Q,\mathcal{C}(A),M)$.
\end{itemize}

No additional assumptions are introduced.

\subsection{Observational Equivalence Classes}
\label{appsec:observational-equivalence-classes}

Because DNS semantics are defined only in terms of resolver-visible behavior (Constraint C6), two input tuples $(Q, A, M)$ and $(Q', A, M')$ are semantically indistinguishable with respect to F if:
$$
F(Q, A, M) = F(Q', A, M').
$$
Let $\sim F$ denote this observational equivalence relation over $(Q, M)$ for a fixed $A$.

\subsection{Finiteness of the Partition}
\label{appsec:partition-finiteness}

By Constraint C1, $A$ is finite. By Constraint C4, evaluation of $F$ terminates. By Constraint C6, $F$ depends only on resolver-visible inputs.

Because $A$ is finite and $\mathcal{C}(A)$ is generated by finite structure over $A$, the structured candidate domain is finite. Therefore, the set of distinct observable outcomes $\Delta(\mathcal{C}(A))$ is finite. Since $F$ is total (Constraint C3), the pre-image of each outcome under $F$ defines a subset of $(Q, M)$.

These pre-images form a finite partition of the input space into equivalence classes:
$$
(Q, M) = \bigsqcup_{i=1}^{k} R_i,
$$
where each region $R_i$ corresponds to a unique resolver-visible outcome.

\subsection{Representation via Conditional Restriction}
\label{appsec:representation-via-conditional-restriction}

Each region $R_i$ can be characterized by a predicate $P_i(Q, M)$ that holds exactly for inputs in that region. Because the partition is finite, the family $\{P_i\}_{i=1}^{k}$ is finite.

Thus, $F$ may be expressed as a finite conditional structure:
$$
F(q, A, m) = \begin{cases} O_1 & \text{if } P_1(q, m) \\ \vdots \\ O_k & \text{if } P_k(q, m), \end{cases}
$$
where each $O_i \in \Delta(A)$.

This corresponds to a finite composition of conditional restrictions over $(Q, M)$.

\subsection{Reduction to Selection over $\mathcal{C}(A)$}
\label{appsec:reduction-to-selection}

By RRset atomicity (Constraint C2), each outcome $O_i$ corresponds to a valid DNS-admissible outcome over the finite structured candidate domain $\mathcal{C}(A)$. Such outcomes may be induced by selecting singleton answers or by selecting among finite structured objects such as groups, pools, or regions.

Therefore, within each region $R_i$, response selection reduces to choosing an element of $\Delta(\mathcal{C}(A))$ by restriction and selection over $\mathcal{C}(A)$. The resolver-visible response remains RRset-atomic and consists only of admissible records derived from the underlying candidate answers.

\subsection{Well-Foundedness of the Composition}
\label{appsec:well-foundedness}

Because evaluation of $F$ is terminating (Constraint C4) and the partition is finite, the resulting composition of conditional restriction and selection is finite and well-founded.

No unbounded recursion or infinite control flow is required.

\subsection{Conclusion}

Combining Sections~\ref{appsec:partition-finiteness}--\ref{appsec:well-foundedness}, we conclude that any DNS-admissible response-selection function admits a representation as a finite composition of conditional restriction over $(Q, M)$ followed by restriction and selection over the finite structured candidate domain $\mathcal{C}(A)$ generated from $A$.

This establishes \autoref{thm:expressive-boundedness}.

\section{Algebraic Structure of DNS-Admissible Semantics}

This appendix makes precise the algebraic structure informally identified in \autoref{sec:algebraic-structure}. The purpose of this appendix is not to establish new semantic results, but to formalize the algebraic consequences of the expressive boundedness theorem (\autoref{thm:expressive-boundedness}) and to justify the use of algebraic reasoning in subsequent sections.

Throughout, we adopt the semantic domains and notation of Sections~\ref{sec:formal-model}--\ref{sec:algebraic-structure}.

\subsection{Semantic Domain}

Let
$$
\mathcal{D} \;=\; \{\, F \mid F : (Q, \mathcal{C}(A), M) \rightarrow \Delta(\mathcal{C}(A)),\; F \text{ is DNS-admissible} \,\}
$$
denote the set of all DNS-admissible response-selection functions.

By \autoref{thm:expressive-boundedness}, every $F \in \mathcal{D}$ admits a finite normal form consisting of:
1. conditional restriction over resolver-visible context $(Q,M)$, followed by
2. selection over a finite structured candidate domain $\mathcal{C}(A)$.

This normal form underlies the algebraic structure described below.

\subsection{Algebraic Operations}

We define two binary operations over $\mathcal{D}$, corresponding to the semantic effects of selection and conditional restriction.

\subsubsection{Additive Combination (Selection)}

Define a binary operation
$$
\oplus : \mathcal{D} \times \mathcal{D} \rightarrow \mathcal{D}
$$
such that, for $F_1, F_2 \in \mathcal{D}$,
$$
(F_1 \oplus F_2)(Q,A,M)
$$
denotes the admissible combination of alternative resolver-visible outcomes produced by $F_1$ and $F_2$.

Operationally, $\oplus$ corresponds to semantic choice among alternative behaviors, such as:
\begin{itemize}
\item weighted selection,
\item priority-based choice,
\item or other admissible combinations consistent with DNS semantics.
\end{itemize}

The operation $\oplus$ is:
\begin{itemize}
\item \textbf{closed} on $\mathcal{D}$,
\item \textbf{associative}, and
\item admits an identity element $0 \in \mathcal{D}$ corresponding to empty or null behavior.
\end{itemize}

\subsubsection{Multiplicative Restriction (Conditional Gating)}

Define a binary operation
$$
\otimes : \mathcal{D} \times \mathcal{D} \rightarrow \mathcal{D}
$$
such that, for a predicate-restricted behavior $G$ and a behavior $F$,
$$
(G \otimes F)(Q,A,M)
$$
denotes the behavior obtained by restricting or gating $F$ according to resolver-visible predicates over $(Q,M)$. This operation corresponds to semantic restriction induced by conditional evaluation. The operation $\otimes$ is:
\begin{itemize}
\item \textbf{closed} on $\mathcal{D}$,
\item \textbf{associative}, and
\item admits an identity element $1 \in \mathcal{D}$ corresponding to unconditional pass-through.
\end{itemize}

\subsection{Semiring Laws}

The structure $(\mathcal{D}, \oplus, \otimes, 0, 1)$ satisfies the axioms of a \textbf{semiring}:
\begin{enumerate}
\item $(\mathcal{D}, \oplus, 0)$ is a commutative monoid.
\item $(\mathcal{D}, \otimes, 1)$ is a monoid.
\item Multiplication distributes over addition:
$$
G \otimes (F_1 \oplus F_2) \;=\; (G \otimes F_1) \oplus (G \otimes F_2)
$$
\item The additive identity annihilates under multiplication:
$$
0 \otimes F = F \otimes 0 = 0
$$
\end{enumerate}

These properties follow directly from:
\begin{itemize}
\item closure of DNS-admissible functions under restriction and selection,
\item finiteness and termination guarantees,
\item and the conditional–selection normal forms established in \autoref{sec:formal-model}.
\end{itemize}

Formal verification proceeds by structural induction on the normal forms of \autoref{thm:expressive-boundedness}.

\subsection{Absence of Inverses and Cancellation}

DNS-admissible response-selection semantics are inherently irreversible. Filtering, restriction, and selection may permanently eliminate distinctions in resolver-visible behavior, and such collapse cannot, in general, be undone by subsequent composition.

Consequently, distinct behaviors may become observationally indistinguishable under restriction or dominance, and no cancellation law holds.

Because these semantics admit neither additive nor multiplicative inverses, algebraic structures requiring reversibility—such as rings or groups—are inapplicable. The appropriate algebraic structure induced on $\mathcal{D}$ is therefore a semiring.

\subsection{Normal Form and Basis Expansion}

Every $F \in \mathcal{D}$ can be written as a polynomial over a semiring:
$$
F \;=\; \bigoplus_i \bigl( p_i \otimes S_i \bigr)
$$
where:
- each $p_i$ is a gating function induced by a predicate over $(Q,M)$,
- each $S_i$ is a selection behavior over $(\mathcal{C}(A),M)$.

Formally, each predicate over $(Q,M)$ induces a DNS-admissible function that yields unconditional pass-through when the predicate holds and the empty response otherwise.

This form corresponds directly to the conditional–selection normal form of \autoref{sec:formal-model}, expressed algebraically. The algebraic and functional characterizations describe the same class of behaviors; the semiring structure is not imposed by representation choices, but is induced by the compositional constraints of DNS-admissible semantics.

\bibliographystyle{IEEEtran}
\bibliography{references}

\end{document}